\documentclass[fleqn,usenatbib]{mnras}

\usepackage{newtxtext,newtxmath}

\usepackage[T1]{fontenc}

\DeclareRobustCommand{\VAN}[3]{#2}
\let\VANthebibliography\thebibliography
\def\thebibliography{\DeclareRobustCommand{\VAN}[3]{##3}\VANthebibliography}

\usepackage{graphicx}	
\usepackage{amsmath}	

\title[Extended radio emission in MS 0735.6+7421 with the VLA]{Extended radio emission in the galaxy cluster MS 0735.6+7421 detected with the Karl G. Jansky Very Large Array}

\author[T. Bégin et al.]{T. Bégin$^{1}$\thanks{E-mail: begin@astro.umontreal.ca},
J. Hlavacek-Larrondo$^{1}$, C.L. Rhea$^{1}$,
M. Gendron-Marsolais$^{2}$,
B. McNamara$^{3,4}$, 
\newauthor
R. J. van Weeren$^{5}$, A. Richard-Laferrière$^{6}$, L. Guité$^{1}$,
M. Prasow-Émond$^{1}$, D. Haggard$^{7,8}$\\
$^{1}$Département de Physique, Université de Montréal, Montréal, QC, Canada\\
$^{2}$European Southern Observatory, Alonso de Córdova 3107, Vitacura, Casilla 19001, Santiago, Chile\\
$^{3}$Department of Physics and Astronomy, University of Waterloo, Waterloo, ON, Canada\\
$^{4}$Perimeter Institute for Theoretical Physics, Waterloo, ON, Canada\\
$^{5}$Leiden Observatory, Leiden University, PO Box 9513, 2300 RA Leiden, The Netherlands\\
$^{6}$Institute of Astronomy, University of Cambridge, Madingley Road, Cambridge, UK\\
$^{7}$Department of Physics, McGill University, 3600 rue University, Montréal, QC, Canada
\\
$^{8}$McGill Space Institute, McGill University, 3550 rue University, Montréal, QC, Canada}

\date{Accepted 2022 November 18. Received 2022 November 16; in original form 2022 February 2}

\pubyear{2022}

\begin{document}
\label{firstpage}
\pagerange{\pageref{firstpage}--\pageref{lastpage}}

\maketitle

\begin{abstract}
MS 0735.6+7421 ($z = 0.216$) is a massive cool core galaxy cluster hosting one of the most powerful active galactic nuclei (AGN) outbursts known. The radio jets of the AGN have carved out an unusually large pair of X-ray cavities, each reaching a diameter of $200$ kpc. This makes MS 0735.6+7421 a unique case to investigate active galactic nuclei feedback processes, as well as other cluster astrophysics at radio wavelengths. We present new low-radio-frequency observations of MS 0735.6+7421 taken with the Karl G. Jansky Very Large Array (VLA): 5 hours of P-band ($224-480$ MHz) and 5 hours of L-band ($1-2$ GHz) observations, both in C configuration. Our VLA P-band ($224-480$ MHz)
observations reveal the presence of a diffuse radio component reaching a scale of $\approx$ $900$ kpc in the direction of the jets and of $\approx$ $500$ kpc in the direction perpendicular to the jets. This component is centered on the cluster core and has a radio power scaled at $1.4$ GHz of $P_{1.4\text{ GHz}} = (4\pm2)\times 10^{24}$ WHz$^{-1}$. Its properties are consistent with those expected from a radio mini-halo as seen in other massive cool core clusters, although it may also be associated with radio plasma that has diffused out of the X-ray cavities, or to a combination of these two hypotheses. Observations at higher spatial resolution are needed to fully characterize the properties and nature of this component. We also suggest that if radio mini-halos originate from jetted activity, we may be witnessing the early stages of this process.
\end{abstract}




\section{Introduction}

Active Galactic Nucleus (AGN) feedback is widely regarded as the main source preventing cooling at the center of cool core clusters of galaxies (e.g. \citealt{mcnamara_mechanical_2012}). This mechanism is powered by accretion onto the central supermassive black hole (SMBH) that then generates relativistic jets emitting synchrotron emission mainly at MHz and also at GHz frequencies. These jets push the intracluster medium (ICM), inflating bubble-like structures identified as X-ray cavities. The precise mechanism of how the energy is transferred from these jets to the ambient ICM remains unclear, although it is likely related to turbulence, shock waves, sound waves, or heating by mixing (\citealt{zhuravleva_turbulent_2014}, \citealt{fabian_deep_2003}, \citealt{fabian_sound_2017}, \citealt{hillel_kinematics_2020}).

MS 0735.6+7421 (hereafter MS0735) is a well-studied example of extreme AGN feedback in a cluster of galaxies (\citealt{donahue_distant_1992}; \citealt{m_donahue_rosat_1995}; \citealt{mcnamara_heating_2005}; \citealt{cohen_resolving_2005}; \citealt{mcnamara_energetic_2009}; \citealt{birzan_radiative_2010}; \citealt{vantyghem_cycling_2014}; \citealt{biava_constraining_2021} and references therein). It is a massive ($M_{500} = (4.5\pm1.2)\times 10^{14} $M$_\odot$, \citealt{hofmann_thermodynamic_2016}), strong cool core cluster ($K_0 = (12.6 \pm 0.6)$ keV cm$^2$, \citealt{vantyghem_cycling_2014}) that was first identified by \cite{donahue_distant_1992} and was subsequently imaged in X-rays by \cite{mcnamara_heating_2005}. It exhibits two extremely large X-ray cavities of approximately $200$ kpc in diameter each, that have been inflated by a pair of relativistic jets among the most powerful known to date in any cluster core (e.g. total jet power of $1.7^{+0.6}_{-0.5}\times10^{46}$ erg s$^{-1}$). These cavities are surrounded by an elliptical shock front (\citealt{vantyghem_cycling_2014}). 

In this system, \cite{mcnamara_heating_2005} was the first to emphasize the spatial correspondence between the radio jets identified with VLA L-band ($1.4$ GHz) data and the X-ray cavities. 
MS0735 was also imaged by \cite{cohen_resolving_2005} who presented VLA observations at $1.425$ GHz in C configuration and at $325$ MHz in B configuration allowing them to compute integrated spectral indexes for the main components of the cluster. The spectral index is defined as $S_\nu \propto \nu^{-\alpha}$, where $S_\nu$ is the flux density, $\nu$ is the frequency and $\alpha$ is the spectral index. We use the same convention for the spectral index for the rest of this paper.

Using new Chandra observations (cumulative exposure time of $477$ ks), \cite{vantyghem_cycling_2014} estimated the required power to inflate the cavities and concluded that the total power exceeds the X-ray bolometric luminosity within the cooling region, thus compensating for the radiative losses of the ICM. They also further studied the shock front that delimits the two large cavity regions and estimated its Mach number to be $M = 1.26^{+0.04}_{-0.09}$. The energy required to inflate these bubbles and to generate the shock front is on the order of $10^{62}$ erg. Moreover, as part of this work, a second pair of smaller X-ray cavities were identified closer to the central AGN. These smaller cavities are characterized by powers significantly smaller than the powers of the large $200$ kpc cavities (around 30 times smaller) suggesting that the AGN power varies over time. The timescale on which these two generations of cavities occurred is nonetheless short enough to prevent the ICM from cooling (i.e. shorter than the cooling time) and thus supports the AGN feedback hypothesis.

More recently, \cite{biava_constraining_2021} presented results of an in-depth spatial radio analysis of MS0735 using new $144$ MHz Low-Frequency Array (LOFAR) data and archival radio data at $235$ MHz, $325$ MHz, $610$ MHz, $1.4$ GHz and $8.5$ GHz. They identified an intermediate-size X-ray cavity located at the southwest of the core and jet-related radio emission that completely fills the large X-ray cavities. The intermediate cavity is of intermediate age compared to the two generations of cavities identified in \cite{vantyghem_cycling_2014}. Their results show very steep spectral indexes across the extent of the radio jets (up to spectral indexes of $\alpha_{610}^{1420} = 3.4\pm0.6$ between $610$ MHz and $1.42$ GHz for the outer lobes of the jets). Based on spectral ageing maps, their work also suggests that the central AGN is active most of the time and is only going through quiescent phases for brief periods.

Given that MS0735 is a massive, cool core cluster, one might argue that in addition to the radio emission produced by the relativistic jets, it is also an ideal candidate for harbouring a radio mini-halo. Indeed, \cite{giacintucci_occurrence_2017} showed that most of the massive cool core clusters host a mini-halo. Radio mini-halos are faint diffuse radio structures often found in cool core clusters, surrounding the central galaxy. They have a scale on
the order of $100-500$ kpc 
(e.g. \citealt{gitti_discovery_2007}, \citealt{giacintucci_new_2013}, \citealt{prasow-emond_multiwavelength_2020}). They are thus smaller than galaxy clusters but are larger than most relativistic jets from the AGN, which are typically on the scale of $30$ kpc (\citealt{von_der_linden_how_2007}), even though some jet-related structures do reach the order of a few hundred kpc (e.g. MS0735). Radio mini-halos also have steep radio spectra with typical spectral indexes of $\alpha>1$.

Because the radiative timescale of the relativistic particles produced by the AGN jets is shorter than the time required to reach the extent of radio mini-halos, these structures have previously been attributed to a new population of relativistic particles produced in situ or to an old population of relativistic particles that have been re-accelerated. The source of relativistic particles responsible for radio mini-halos has been historically debated between the hadronic model and the turbulent model. In the hadronic model, the electrons are produced as a result of the hadronic interaction between cosmic ray (CR) protons and ICM protons (\citealt{pfrommer_constraining_2004}; \citealt{fujita_nonthermal_2007}; \citealt{zandanel_physics_2014}; \citealt{ignesti_radio_2020}). However, this model would imply extended emission of $\gamma$-rays which have not been detected up to this day (see \citealt{ahnen_deep_2016}). In the turbulent model, seed electrons are re-accelerated by ICM turbulence induced by phenomena such as gas sloshing (see \citealt{zuhone_turbulence_2013} for simulation results). This hypothesis is the most accepted, partly because most radio mini-halos are bounded by cold fronts (\citealt{giacintucci_radio_2011}) which are a typical sign of gas sloshing (see \citealt{markevitch_shocks_2007}).

Some studies have previously suggested that relativistic seed particles responsible for radio emission of radio mini-halos could originate from the central AGN (e.g. \citealt{fujita_nonthermal_2007}).
Moreover, other studies suggested that the central AGN might be the source of the turbulence that drives the re-acceleration of particles in radio mini-halos (e.g. \citealt{bravi_radio_2016}). 
This result was further supported by the discovery of strong correlations between the radio power of radio mini-halos and the brightest cluster galaxy (BCG) steep radio power component, and between the radio mini-halo radio power and the X-ray cavity power which suggests a link between the feedback processes of the central AGN and radio mini-halos (\citealt{richard-laferriere_relation_2020}). Additional high spatial resolution and high sensitivity radio observations of clusters exhibiting radio mini-halos and hosting extreme AGN feedback are required to test these hypotheses. As previously pointed out, MS0735 is a target of choice for such an investigation because it exhibits uniquely large relativistic jets characteristic of extreme central AGN feedback, yet no study has reported the detection of a radio mini-halo in this cluster. 

In this paper, we present a radio analysis of MS0735 using new, deep, radio data acquired using the VLA in P-band ($224-480$ MHz) and L-band ($1-2$ GHz), both in C configuration. Our data is probing spatial scales suitable for the detection of a radio mini-halo, if present. To complement our analysis, we used X-ray data from the Chandra observatory archival dataset. Section 2 presents the radio and X-ray observations and the data reduction process. Section 3 presents the resulting images and their analysis. In section 4, we discuss the results and their implications. Finally, in section 5, we present our main conclusions and the objectives for future studies related to MS0735.

Throughout the paper, we use  $H_0 = 70$ km s$^{-1}$ Mpc$^{-1}$, $\Omega_m = 0.3$ and $\Omega_\Lambda = 0.7$ assuming a $\Lambda$CDM cosmological model and $z = 0.216$ as the cluster's redshift ($1$ arcsec $=3.5$ kpc). Errors are 1$\sigma$ unless otherwise stated. 


\section{Observations and data reduction}
\label{Observations}

\subsection{New VLA observations}
We obtained a total of 5 hours in P-band ($224-480$ MHz) and 5 hours in L-band ($1-2$ GHz) using the VLA (project number 18B-171, principal investigator J. Hlavacek-Larrondo). Both of these observations were taken in C configuration, on December 5th and December 6th of 2018 respectively.

\begin{table*}
	\caption{New radio observations summary}
	\label{table}
	\begin{tabular}{ccccccccc} 
		\hline
		ID & Band & Frequency range & Robust value & Beam size & Sensitivity & Peak brightness & Fig. number\\
		\hline
		1 & P & 224 - 480 MHz & -2.0 & 27 $\times$ 45 arcsec & 0.98 mJy beam$^{-1}$ & 176 mJy beam$^{-1}$ & 2 (top middle)\\
		2 & P & 224 - 480 MHz & 2.0 & 59 $\times$ 75 arcsec & 0.50 mJy beam$^{-1}$ & 314 mJy beam$^{-1}$ & 2 (bottom middle), 5 (left)\\
		3 & L & 1 - 2 GHz & -2.0 & 6.6 $\times$ 11.4 arcsec & 0.0165 mJy beam$^{-1}$ & 5.00 mJy beam$^{-1}$ & 2 (top right)\\
            4 & L & 1 - 2 GHz & 2.0 & 13.5 $\times$ 19.2 arcsec & 0.0170 mJy beam$^{-1}$ & 7.92 mJy beam$^{-1}$ & 1, 2 (bottom right), 4\\
		\hline
	\end{tabular}
\end{table*}

The observations were taken with 27 operational antennas. Within the observation periods, no anomaly was reported by the operator log. The VLA L-band ($1-2$ GHz) dataset consists of 45 scans, from which three scans are of 3C147, one at the beginning, one in the middle, and one at the end (for flux and bandpass calibration), 10 scans are of J0841+7053 (for phase calibration) and the rest of the scans are of MS0735. The VLA L-band ($1-2$ GHz) receiver has 16 spectral windows (bandwidth of $64$ MHz), each subdivided into 64 channels (bandwidth of $1$ MHz). The VLA P-band ($224-480$ MHz) dataset consists of 45 scans, from which three scans are of 3C147, one at the beginning, one in the middle, and one at the end (for flux and bandpass calibration), ten scans are of J0939+8315 (for phase calibration) and the rest of the scans are of MS0735. The VLA P-band ($224-480$ MHz) receiver has 16 spectral windows (bandwidth of $16$ MHz), each subdivided into 128 channels (bandwidth of $125$ kHz). 

The data reduction of both of the datasets was done using the Common Astronomy Software Applications - CASA (\citealt{mcmullin_casa_2007}) version 5.6.2-2 following the procedure described in the low-frequency VLA P-band ($224-480$ MHz) tutorial found on the National Radio Astronomy Observatory (NRAO) website: \url{https://casaguides.nrao.edu/index.php?title=Main_Page}. A pipeline was specifically developed to account for the strong presence of radio frequency interference (RFI) at these low-frequency. The steps of the data reduction process are detailed in the following paragraphs.

First, malfunctioning antennas and scans impacted by evident RFI were flagged according to the visual inspection of the datasets. Calibrations for the antenna position, for the ionospheric total electron content (TEC), and for the requantizer gains at the inputs of the WIDAR correlator were derived and applied. The data were then Hanning smoothed utilizing the \texttt{HANNINGSMOOTH} task. Initial automatic flagging was then conducted on the calibrators and on the target to remove the most apparent RFI using the \texttt{FLAGDATA} task in \texttt{TFCROP} mode. A preliminary bandpass calibration was conducted and applied to the data using the \texttt{FLAGDATA} task in \texttt{RFLAG} mode to reduce even more the missed RFI. After these initial flagging steps, calibration was conducted. For both L-band ($1-2$ GHz) and P-band ($224-480$ MHz) datasets, 3C147 was used as the bandpass and flux calibrator while the phase calibrators were respectively J0841+7053 and J0939+8315. The data were calibrated using the standard calibration tasks (\texttt{SETJY}, \texttt{GAINCAL}, \texttt{BANDPASS}, \texttt{SMOOTHCAL}) and we proceeded to visually inspect each calibration table produced with the \texttt{PLOTMS} task, thus allowing us to identify and flag any obvious outliers. After applying the derived calibrations using the \texttt{APPLYCAL} task, the corrected data were split using the \texttt{SPLIT} task and a final weight down of the RFI was applied using the \texttt{STATWT} task. The deconvolution was performed with a self-calibration method using the task \texttt{TCLEAN} and two subsequent iterative cycles of self-calibration were applied to produce incremental gain phase corrections. The self-calibration step used the main target MS0735 instead of a calibration source to refine the final calibration of the image. For both bands, the deconvolution was done in interactive mode to progressively build a mask indicating emission regions to the \texttt{TCLEAN} algorithm. The parameters for the deconvolution by the \texttt{TCLEAN} task were carefully chosen. A multi-term and multi-frequency synthesis-imaging algorithm (MS-MFS, \citealt{rau_multi-scale_2011}) was used. The different scales used were chosen to account for the point sources, the jets, and a radio mini-halo typical size. We used 2 Taylor's coefficients assuming a linear spectrum for the source. We used the W-projection algorithm to correct for the widefield non-coplanar baselines effect (\citealt{cornwell_non-coplanar_2008}). We used 128 and 256 w-planes for the L-band ($1-2$ GHz) and P-band ($224-480$ MHz) respectively, to get rid of artifacts resulting from the effect of the non-coplanar baseline while minimizing computing costs. We used a Briggs weighting algorithm and set the pblimit parameter to -0.02 to prevent masking the edges of the image. We run $100000$ iterations for each \texttt{TCLEAN}. For both bands, we tested the three main Brigg's robustness parameters of -2.0, 0.5, and 2.0 to either maximize the spatial resolution (robust of -2.0) or maximize the diffuse emission detection (robust of 2.0).

We also produced a compact sources image with the VLA L-band ($1-2$ GHz) data by applying a \textit{uvrange} cutoff ($>\text{10k}\lambda$) in the \texttt{TCLEAN} task. This \textit{uvrange} cutoff corresponds to angular scales of 20 arcsec which is approximately twice the synthesized beam size. This allowed producing an image utilizing only the longest baselines of the interferometer, thus allowing to image compact sources. 

The size of the VLA L-band ($1-2$ GHz) images (1260 pixels $\times$ 1260 pixels \textasciitilde 1.4$^\circ \times$ 1.4$^\circ$, while the full width at half power of the field of view is around 0.5$^\circ$) and of the P-band ($224-480$ MHz) images (1260 pixels $\times$ 1260 pixels \textasciitilde 5.6$^\circ \times$ 5.6$^\circ$, while the full width at half power of the field of view is around 2.4$^\circ$) were chosen to be large enough to include all bright sources in the vicinity of MS0735. 
The images presented in the paper were cropped to show the central part of interest of the cluster.

For the VLA L-band ($1-2$ GHz) data reduction, even after testing every antenna as the reference antenna for the calibration, we ended up flagging approximately $60$ per cent of the data due to the large amount of RFI. For the P-band ($224-480$ MHz), we flagged approximately $65$ per cent of the data. For both bands, we reached the expected rms noise computed using the online VLA Exposure Calculator: \url{https://obs.vla.nrao.edu/ect/}. A summary of our radio observations is presented in table \ref{table}.

\begin{figure}
    \centering
    \includegraphics[scale=0.275]{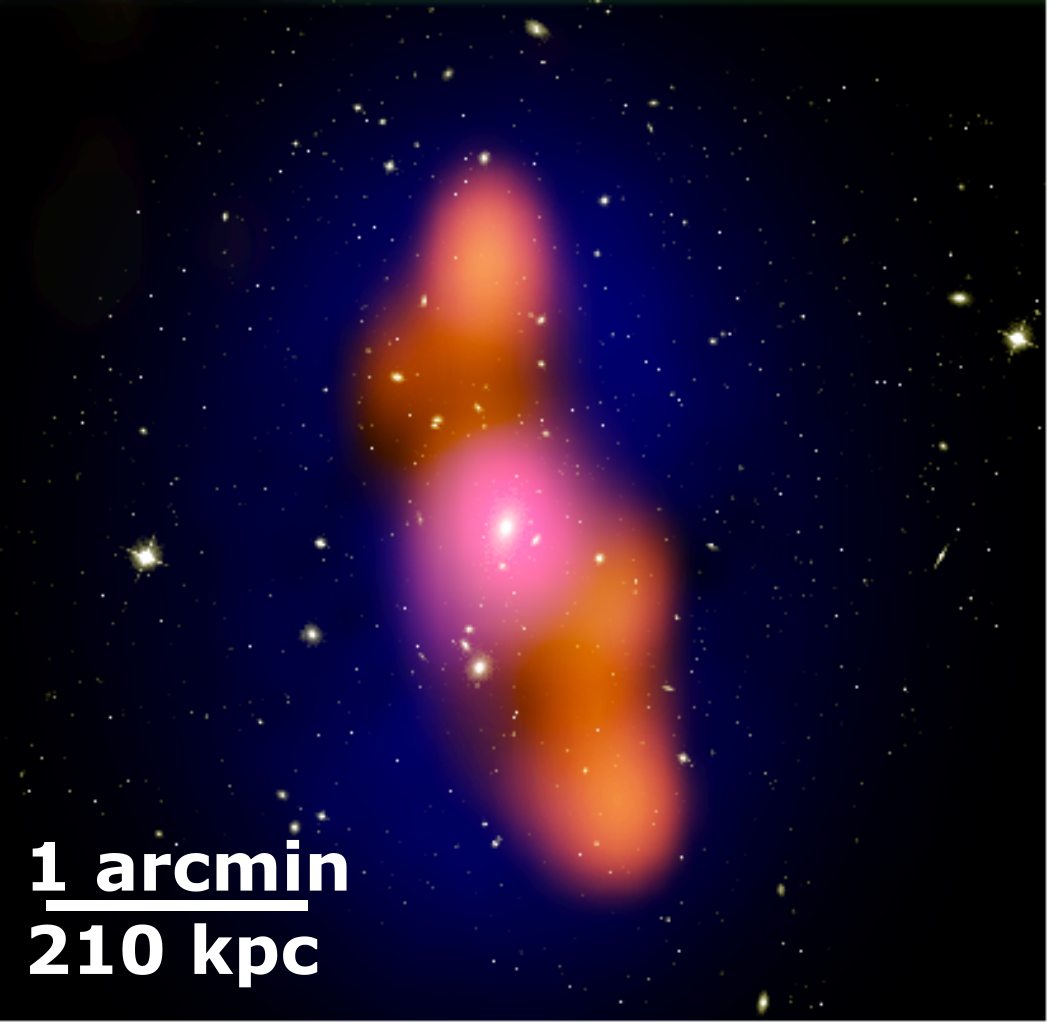}
    \caption{Composite image of the cluster's central $3.7 \times 3.7$ arcmin. The image shows a combination of the unsharp-masked X-ray soft-band ($0.5-2.0$ keV) emission (blue), the optical emission (white), and the new VLA L-band ($1-2$ GHz) radio emission (orange).}
    \label{composite}
\end{figure}

\begin{figure*}
    \centering
    \includegraphics[scale=0.5175]{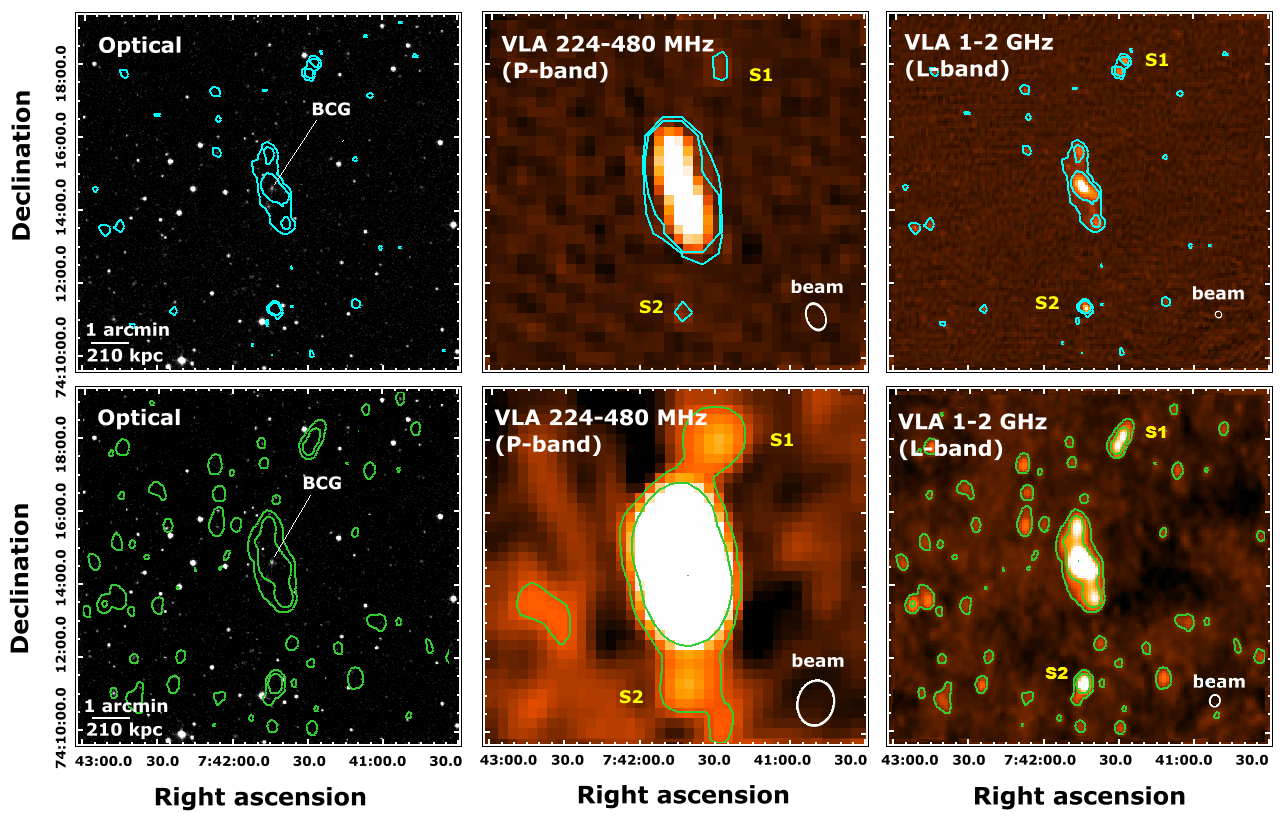}
    \caption{All images are on the same scale and cover a region of approximately $10 \times 10$ arcmin.  \textbf{Top left:} Samuel Oschin Telescope optical image of the cluster in red filter with VLA L-band (1-2 GHz) contours maximizing the spatial resolution (radio observation ID = 3, see table \ref{table}). \textbf{Top middle:} VLA P-band ($224-480$ MHz) maximizing spatial resolution (radio observation ID = 1, see table \ref{table}). Contour levels are drawn at [1, 3, 59]$\times 4 \sigma_{\text{rms}}$ (3 levels in total). \textbf{Top right:} VLA L-band ($1-2$ GHz) maximizing spatial resolution (radio observation ID = 3, see table \ref{table}). Contour levels are drawn at [1, 4, 101]$\times 3 \sigma_{\text{rms}}$ (3 levels in total). \textbf{Bottom left:} Same optical image as top-left but with VLA L-band ($1-2$ GHz) contours maximizing diffuse emission detection (radio observation ID = 4, see table \ref{table}).  \textbf{Bottom middle:} VLA P-band ($224-480$ MHz) maximizing diffuse emission detection (radio observation ID = 2, see table \ref{table}). Contour levels are drawn at [1, 7, 209]$\times 3 \sigma_{\text{rms}}$ (3 levels in total). \textbf{Bottom right:} VLA L-band ($1-2$ GHz) maximizing diffuse emission detection (radio observation ID = 4, see table \ref{table}). Contour levels are drawn at [1, 6, 155]$\times 3 \sigma_{\text{rms}}$ (3 levels in total).}
    \label{nouvelle_fig2}
\end{figure*}

\begin{figure}
    \centering
    \includegraphics[scale=0.65]{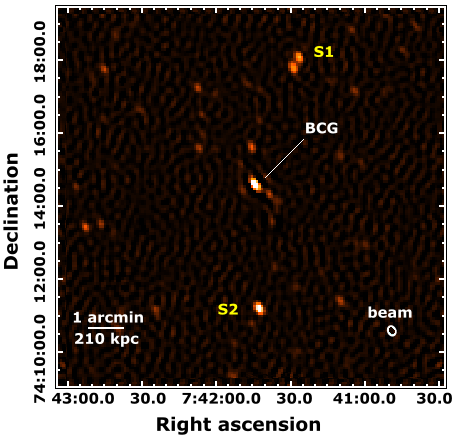}
    \caption{Compact source VLA L-band ($1-2$ GHz) image of MS0735. The field of view is approximately $10\times 10$ arcmin. The noise level in the vicinity of the cluster is $\sigma_\text{rms} = 0.025$ mJy beam$^{-1}$, the peak value is $3.40$ mJy beam$^{-1}$ and the beam size is $7.4$ $ \times$ $12.0$ arcsec.}
    \label{PS}
\end{figure}

\subsection{Chandra observations}
We used archival Chandra observations, all taken with the Advanced CCD Imaging Spectrometer (ACIS) in VFAINT mode. The observations were taken in June 2009. There are 7 observations in total: 10468 (46ks), 10469 (97ks), 10470 (150ks), 10471 (20ks), 10822 (79ks), 10918 (68ks), 10922 (38ks) for a total of 498 ks.  Each ObsID was cleaned using the \texttt{CIAO} software (\texttt{CIAO v4.10} and \texttt{CALDB v4.7.8}). First, the level one event files were processed for background flare events using the standard \texttt{lc\_sigma\_clip} routine with $\sigma$ set to 3. Then, \texttt{chandra\_repro} was applied to the unprocessed data with \texttt{vfaint} set to \texttt{true}. Afterward, we created blanksky background files for each observation with \texttt{blanksky}. Using \texttt{blanksky$\_$image}, we created background-subtracted, exposure-corrected images of each observation. Finally, the \texttt{merge\_obs} command was used to merge the observations. Five point sources were visually noted and removed from the final image. We used the \texttt{dmfilth} command and a \texttt{Poisson} distribution to fill in these regions.  

The data were spectroscopically fit using \texttt{Xspec v12.10.1}, \texttt{Sherpa  v1}, and \texttt{python v3.5}; moreover, the observations were fit simultaneously over $0.7-8.0$ keV. We employed an absorbed thermal model to describe the ICM emission: \texttt{PHABS*APEC}. We adopted a column density of $3.29\times10^{20}$~cm$^{-2}$ (\citealt{kalberla_leidenargentinebonn_2005}). The background spectrum was subtracted from the region's extracted spectrum prior to applying the fit. We used a standard reduced $\chi^2$ fit statistic and Levenberg-Marquardt optimization algorithm as implemented in \texttt{Sherpa}. The regions were created using a Weighted Voronoi Tessellation (WVT) algorithm implemented in \texttt{python} applied to the 0.7 -- 8.0 keV background-subtracted merged X-ray image; the implementation can be found at: \url{https://github.com/XtraAstronomy/AstronomyTools} (\citealt{rhea_x-tra_2020}). The WVT algorithm bins regions in geometrically unbiased groups meeting a pre-determined signal-to-noise requirement. We set the target signal-to-noise to 80. The fitting process was completed in three stages: 
\begin{enumerate}
    \item Freeze the metallicity to $0.3$Z$_\odot$ and fit the normalization and temperature;
    \item Freeze the temperature and thaw the metallicity, redo the fit;
    \item Thaw the temperature and fit with both metallicity and temperature free.
\end{enumerate}
In this manner, the fits were able to first constrain the temperature, the metallicity, and then ensure that the estimates are stable. Finally, we calculated the luminosity in the $0.1-2.4$ keV band using \texttt{Sherpa}'s \texttt{sample\_flux} command.

\section{Results}
\label{Results}

Figure \ref{composite} shows a composite image of MS0735. In blue, we show the Chandra soft-band ($0.5-2.0$ keV) X-ray image after an unsharp-mask was applied. In white, we show the optical image taken with the Hubble Space Telescope's Advanced Camera for Surveys (ACS) through the F850LP filter. Finally, in orange, we show the VLA L-band ($1-2$ GHz) image of the bottom right panel of figure \ref{nouvelle_fig2}. This image highlights the correspondance between the radio emission and the X-ray cavities. 

The top panels of figure \ref{nouvelle_fig2} show the VLA L-band ($1-2$ GHz) and VLA P-band ($224-480$ MHz) images maximizing the spatial resolution (robust parameter of -2.0 in the \texttt{TCLEAN} task) as well as an optical image that was taken with the Samuel Oschin Telescope in the red filter. In the radio images, we can identify two bright compact sources located at the northwest and the south of the jets of MS0735. We respectively call these sources S1 and S2 (see yellow identification in figure \ref{nouvelle_fig2}). The VLA L-band ($1-2$ GHz) image in the top right panel shows the structure of the jets while the VLA P-band ($224-480$ MHz) image in the top middle panel only shows the general large-scale structure due to its lower spatial resolution in comparison with the L-band ($1-2$ GHz). 
The bottom panels of figure \ref{nouvelle_fig2} show the VLA L-band ($1-2$ GHz) and VLA P-band ($224-480$ MHz) images maximizing the diffuse emission detection (robust parameter of 2.0 in the \texttt{TCLEAN} task) as well as the same Samuel Oschin Telescope red filter optical image. The bottom right VLA L-band ($1-2$ GHz) image has a rms of $0.017$ mJy beam$^{-1}$, a peak value of $7.92$ mJy beam$^{-1}$, and a beam size of $13.5$ $\times $ $19.2$ arcsec\footnote{all beam sizes in the manuscript refer to minor and major axis full width respectively.}, while the bottom middle VLA P-band ($224-480$ MHz) image has a rms of $0.5$ mJy beam$^{-1}$, a peak value of $314$ mJy beam$^{-1}$, and a beam size of $59$ $\times $ $75$ arcsec. 

In figure \ref{PS}, we
show a radio compact source image in the vicinity of MS0735 (produced with a \textit{uvrange} cutoff of >10k$\lambda$ in the \texttt{TCLEAN} task). We attempted to image only the extended emission by utilizing the \texttt{UVSUB} task: we first subtracted the VLA L-band ($1-2$ GHz) compact source image from the initial dataset and then re-imaged. However, the rms of the final image was insufficient to infer the requisite physical information.

\begin{figure*}
    \centering
    \includegraphics[scale=0.505]{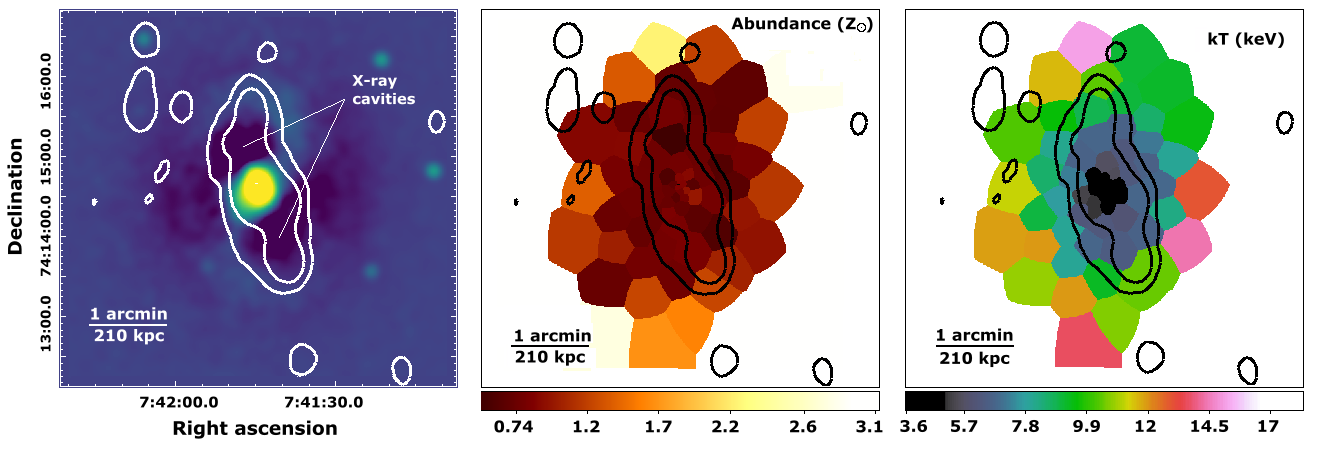}
    \caption{All images are on the same scale and cover a region of approximately $5 \times 5$ arcmin. \textbf{Left:} Unsharp-masked Chandra soft-band ($0.5-2.0$ keV) image. A difference of Gaussians was applied (standard deviations of 6 and 60 pixels respectively) to improve the contrast in order to better emphasize the X-ray cavities. The X-ray image is overlaid with the VLA L-band ($1-2$ GHz) radio contours maximizing diffuse emission detection (radio observation ID = 4, see table \ref{table}). \textbf{Middle:} Abundance map in units of Z$_\odot$ (linear scale) with the same VLA L-band ($1-2$ GHz) radio contours overlaid. \textbf{Right:} Temperature map in units of keV (linear scale) with the same VLA L-band ($1-2$ GHz) radio contours overlaid.}
    \label{soft_xrays_Lband_contours}   
\end{figure*}

In figure \ref{soft_xrays_Lband_contours}, we compare the VLA L-band ($1-2$ GHz) radio contours to the Chandra soft-band X-ray ($0.5-2.0$ keV) unsharp-masked image. The image demonstrates a spatial correlation between the radio jets and the ICM emission, but the radio emission goes beyond the X-ray cavities. We also present an abundance map and a temperature map both with the VLA L-band ($1-2$ GHz) radio contours overlaid. The temperature map shows an increase from a temperature of $T_\text{center} = 3$ keV to the center to a temperature of $T_\text{edge} = 11$ keV to the edge.  The cooler gas in the central regions of the cluster is slightly elongated in the direction of the jets as was pointed out in \cite{vantyghem_cycling_2014}. This effect is most likely due to the centralized cold gas being dragged out by the jets. The abundance map shows a central value of $Z_\text{central} = 0.8$ Z$_\odot$. Along the axis of the jets, the abundance decreases from the center until it reaches a value of approximately $0.35$ Z$_\odot$. This is consistent with work from \cite{grandi_metal_2009} that presents this central enrichment as a result of supernovae in the BCG. The abundances then increase back to values of approximately $Z_\text{edge} = 1.2$ Z$_\odot$ at about $300$ kpc from the center of the cluster. This is consistent with results presented in \cite{vantyghem_cycling_2014} and with the work done by \cite{kirkpatrick_direct_2009} and \cite{simionescu_chemical_2009} which argue that this increase in abundance at a large radius from the center is due to uplifting of the metals by the jets. 

\begin{figure*}
    \centering
    \includegraphics[scale=0.67]{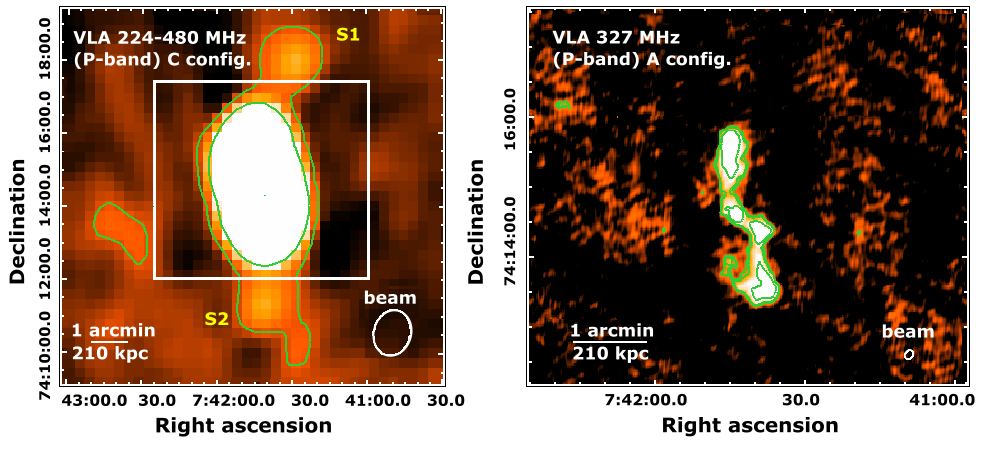}
    \caption{Comparison between the previously published VLA P-band ($327$ MHz) A configuration image from \protect \cite{mcnamara_energetic_2009} and our new VLA P-band ($224-480$ MHz) C configuration image. \textbf{Left:} VLA P-band ($224-480$ MHz) maximizing diffuse emission detection (radio observation ID = 2, see table \ref{table}). The white square region corresponds to the field of view covered in the right panel. \textbf{Right:} previously published VLA P-band ($327$ MHz) A configuration image obtained from \protect \cite{mcnamara_energetic_2009}. The noise level in the vicinity of the cluster is $\sigma_\text{rms} = 1.4$ mJy beam$^{-1}$, the peak value is $46.4$ mJy beam$^{-1}$ and the beam size is $5.7$ $ \times $ $7.6$ arcsec. Contour levels are drawn at [1, 1.3, 2.8, 11]$\times 3 \sigma_{\text{rms}}$ (4 levels in total).}
    \label{comparison_old_new_Pband}
\end{figure*}

Figure \ref{comparison_old_new_Pband} shows the comparison between the previously published VLA P-band ($327$ MHz) A configuration image from \cite{birzan_radiative_2010} and our new VLA P-band ($224-480$ MHz) C configuration image taken from figure \ref{nouvelle_fig2} bottom middle panel. The previously published VLA P-band ($327$ MHz) A configuration image shows a higher spatial resolution with its beam size of $5.7$ $\times$ $7.6$ arcsec. This image allows a precise calculation of the flux density associated with the jets and the central point source (see section \ref{new_diffuse} for the flux density comparison between both VLA P-band images).

\begin{figure}
    \centering
    \includegraphics[scale=0.215]{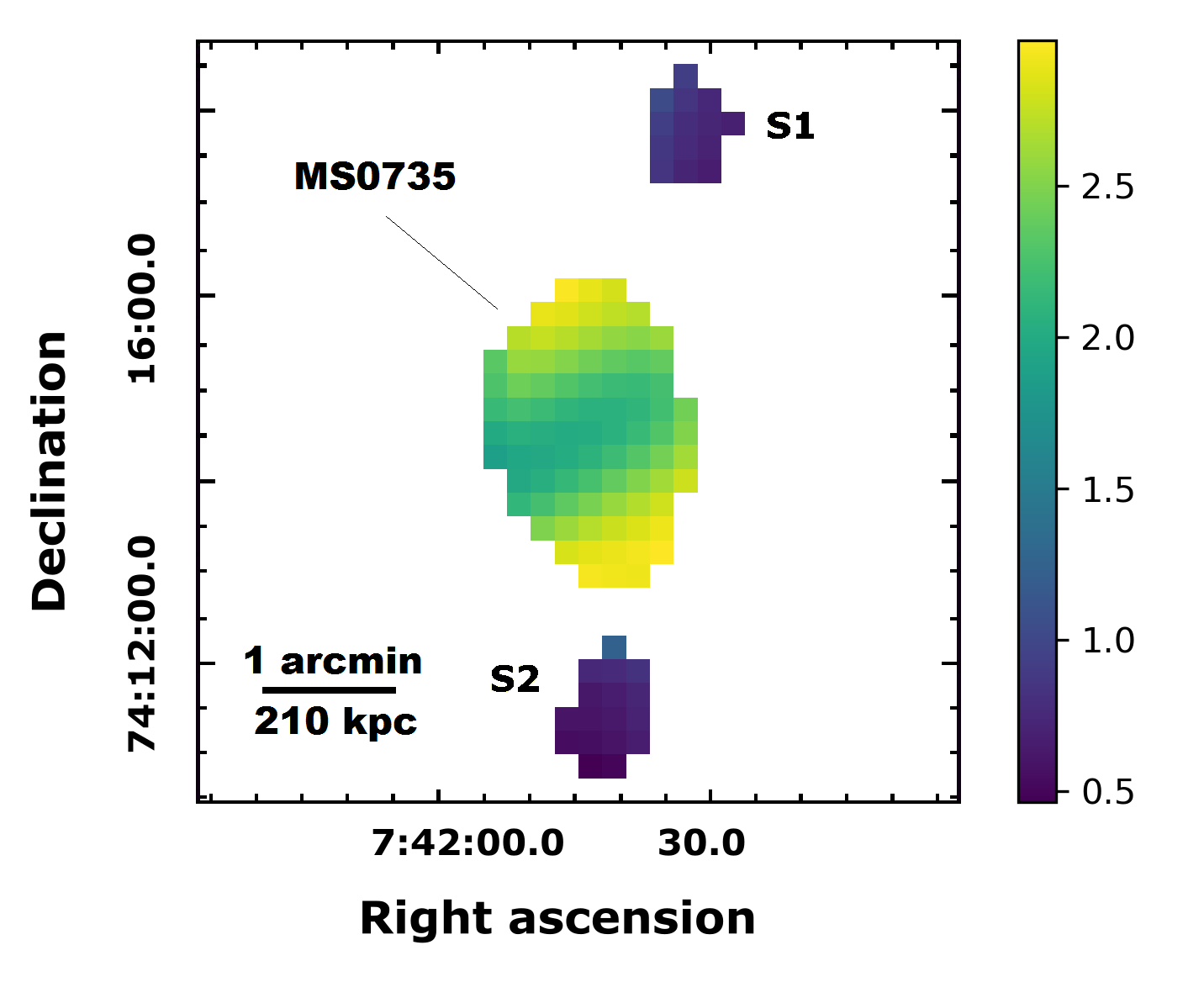}
    \caption{Spectral index map of the radio emission of MS0735 and its surroundings between the new VLA L-band ($1-2$ GHz) and VLA P-band ($224-480$ MHz) images. Both islets located at the northwest and at the south of MS0735 correspond to the position of the sources S1 and S2 that were identified in figure \ref{nouvelle_fig2}.}
    \label{SIM}
\end{figure}

Figure \ref{SIM} shows the spectral index map that was produced by comparing our new VLA L-band ($1-2$ GHz) and VLA P-band ($224-480$ MHz) image's flux densities pixel by pixel. To produce the map, we used the images maximizing diffuse emission detection in both bands (radio observation IDs = 2 and 4, see table \ref{table}). The VLA L-band ($1-2$ GHz) image had to be scaled to the spatial resolution of the VLA P-band ($224-480$ MHz) image. The scaling of spatial resolution was done by re-imaging the VLA L-band ($1-2$ GHz) dataset with the \texttt{TCLEAN} task, but this time by forcing the beam size and the pixel size to match the VLA P-band ($224-480$ MHz) image. We then used the task \texttt{IMREGRID} to regrid the VLA L-band ($1-2$ GHz) image onto the VLA P-band ($224-480$ MHz) one. The colored regions are where the flux density was significant (3 times the average rms noise in the vicinity of the cluster) in both the VLA L-band ($1-2$ GHz) and VLA P-band ($224-480$ MHz) images. The map has three main components: the sources S1 and S2 (as identified in figure \ref{nouvelle_fig2}) and MS0735 which is composed of its central AGN emission, jet emission, and extended diffuse emission. Compact sources S1 and S2 have spectral indexes contained in the interval $0.5-1.0$. The emission of MS0735 has a central spectral index of $\alpha_{\text{center}} = 2.0$ with a steepening gradient towards the outskirt to values up to $\alpha_{\text{outskirt}} = 3.0$. 

\begin{figure*}
    \centering
    \includegraphics[scale=0.182]{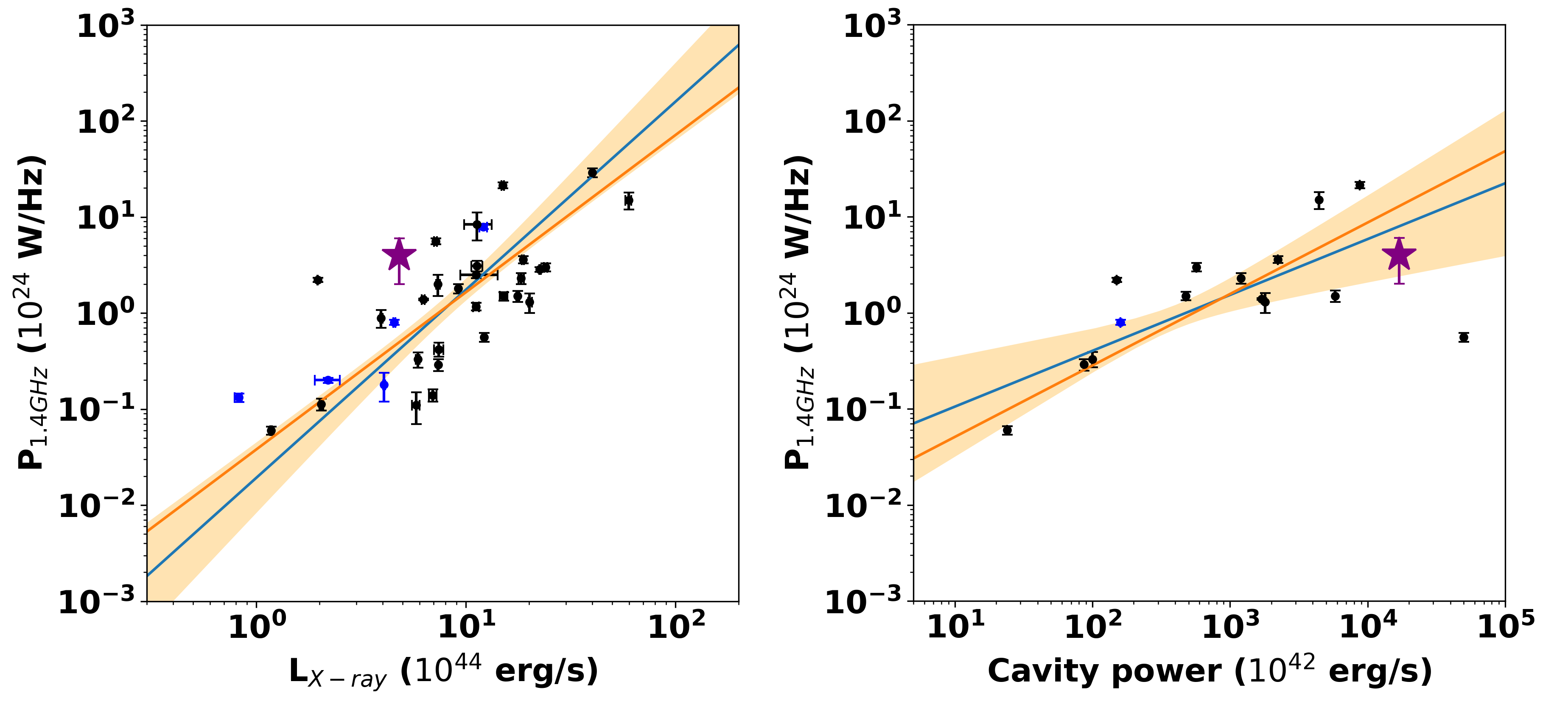}
    \caption{Comparison between the properties of 33 radio mini-halos listed in \protect \cite{richard-laferriere_relation_2020} (blue are candidate or uncertain radio mini-halos and black are confirmed radio mini-halos as identified in the publication)
    and the properties of the diffuse emission detected in the current study in the MS0735 system (purple star). Note that the purple point assumes that the detected excess emission is only attributable to a mini-halo structure (see section \ref{combination}). The best-fit lines using the BCES-orthogonal (blue lines) and the BCES-bisector (orange lines) methods are displayed. The $95$ per cent confidence regions of the best-fit relation for the BCES-orthogonal method are also shown as the orange regions in both panels. In both panels, the best-fit lines, as well as the $95$ per cent confidence regions, were fitted using the 33 radio mini-halos presented in \protect \cite{richard-laferriere_relation_2020} (excluding the purple point for MS0735). \textbf{Left:} radio power ($P_{1.4\text{ GHz}}$) and X-ray luminosity inside a radius of 600 kpc ($L_{\text{X-ray}}$) plane. \textbf{Right:} radio power ($P_{1.4\text{GHz}}$) and X-ray cavity power plane.}
    \label{correlations}
\end{figure*}

\section{Discussion}
\subsection{Previous radio imaging of MS0735}

Previous VLA observations of MS0735 were presented in \cite{mcnamara_heating_2005}. These data were taken in L-band ($1.4$ GHz) C configuration with the non-upgraded VLA (before the 2011 electronic hardware upgrade). The detected radio emission is associated with the radio jets generated by the AGN and is enclosed in the X-ray cavities. No diffuse radio emission was detected as part of this first radio investigation of MS0735. 

\cite{cohen_resolving_2005} further presented VLA observations of MS0735 at $1.425$ GHz in C configuration and $325$ MHz in B configuration. In this study, they computed an integrated spectral index of $\alpha_{325}^{1400} = 1.54 \pm 0.04$ for the core (stating this value can be contaminated by radio lobes) and integrated spectral indexes of $\alpha_{325}^{1400} = 3.17 \pm 0.05$ and $\alpha_{325}^{1400} = 3.13 \pm 0.05$ for the north and south lobes respectively. They also computed a spectral index with the total flux of the structure (core and lobes) between $325$ MHz and $1.425$ GHz and found a value of $\alpha_{325}^{1.425} = 2.45 \pm 0.04$. 

Additionally, VLA observations were presented in \cite{mcnamara_energetic_2009}. These data were taken in VLA P-band ($327$ MHz) A configuration (thus allowing to reach a higher spatial resolution) and are shown in figure \ref{comparison_old_new_Pband} along with our new VLA P-band ($224-480$ MHz) C configuration image. This previous $327$ MHz observation shows that the jets are emerging from the nucleus oriented to the northeast and the southwest. The jets then change direction towards the north and the south and expand in lobe structures due to the resistance offered by the ICM (\citealt{mcnamara_energetic_2009}). The jets are enclosed in the X-ray cavities but do not completely fill them. No diffuse extended radio emission was detected as part of this investigation.

\cite{birzan_radiative_2010} presented an in-depth analysis of multiple radio galaxies at the cores of cooling clusters. For MS0735, they computed a total spectral index (core and lobes) of $\alpha_{327}^{1400} = 2.48 \pm 0.04$ which is supported by the results presented in \cite{cohen_resolving_2005}.

Finally, \cite{vantyghem_cycling_2014} used $pV$, where $p$ is the cavity's pressure and $V$ is its volume, as an estimate of the energy required to inflate the cavities. They used a mean age for the cavities by averaging the sound crossing time, the refill time, and the buoyancy time. The power required to inflate the cavities is $9_{-4}^{+5} \times 10^{45}$ erg s$^{-1}$ and $(8\pm4) \times 10^{45}$ erg s$^{-1}$ for the north and south cavities respectively. This corresponds to a total power of $1.7_{-0.5}^{+0.6}\times10^{46}$ erg s$^{-1}$ which exceeds the X-ray bolometric luminosity within the cooling radius of $2.6\times 10
^{44}$ erg s$^{-1}$ thus compensating for the radiative losses. They also identified a second pair of smaller cavities closer to the central AGN and the power required to inflate these rejuvenated cavities is $2.8_{-1.3}^{+1.6} \times 10^{44}$ erg s$^{-1}$ and $2.4_{-1.3}^{+2.0}\times 10^{44}$ erg s$^{-1}$ for the north and south cavities respectively. This corresponds to a total power of $5.2_{-1.8}^{+2.6}\times10^{44}$ erg s$^{-1}$. Moreover, this detailed X-ray analysis of MS0735 did not reveal a cold front, but instead revealed a shock front surrounding the X-ray cavities.

Overall, these studies suggest that the radio emission detected thus far in MS0735 is entirely due to the relativistic jets originating from the central AGN. Deeper observations were therefore needed to determine if there is evidence of fainter emission, as would be expected from the presence of structures such as a radio mini-halo. 

\subsubsection{144 MHz LOFAR observations}

More recently, \cite{biava_constraining_2021} presented new LOFAR observations at $144$ MHz of MS0735. These observations show a greater spatial correspondence between the morphologies of the jet-related emission and of the X-ray cavities. Indeed, the jet-related emission completely fills the cavities. Moreover, this study was able to provide detailed spectral index maps: they computed a spectral index of $\alpha_{144}^{610} = 1.5 \pm 0.1$ for the core which confirms that the central emission corresponds to a superposition between the core and the lobes' emission. A gradient is also present in their spectral index maps: in the extreme edge of the lobes, they found a spectral index of $\alpha_{144}^{610} = 2.3 \pm 0.1$, this spectral index then steepens going toward the core, reaching a value of $\alpha_{144}^{610} = 2.9 \pm 0.3$ in the outer lobes and flattens again until it reaches a value of $\alpha_{144}^{610} = 2.1 \pm 0.1$ in the intermediate lobe. If there was only one phase of jet activity, the flatter spectral index should have been at the extreme edge of the jets, where the particles are last accelerated. This supports the idea that the radio emission is the result of multiple generations of AGN activity. They also presented a spectral index map between $610$ MHz and $1.42$ GHz which showed spectral steepening at high frequencies compared to low frequencies. They found a spectral index of $\alpha_{610}^{1420} = 3.4 \pm 0.6$ in the outer lobes, a value of $\alpha_{144}^{610} = 2.8 \pm 0.3$ in the intermediate lobe and a value of $\alpha_{144}^{610} = 1.4\pm 0.1$ in the core.
The observed spectral curvature (steepening of spectral indexes at higher frequencies) is characteristic of an aged population of electrons. They also provided spectral ageing maps allowing to investigate the rejuvenation timescales of the central AGN. As part of this study, they identified a wider than previously detected radio emission that they attributed to the jets. This result is well supported by the correspondence between the $144$ MHz radio emission morphology and the X-ray cavities, by the spectral index maps they derived, and by their re-imaging of their data at lower spatial resolution where they do not detect more extended emission. The flux density they measured for the total structure (lobes and central core) at $144$ MHz is $S_{144\text{MHz}} = (4.7\pm0.5)$ Jy. 

\subsection{Newly-discovered extended radio emission with the VLA}
\label{new_diffuse}

\subsubsection{VLA L-band ($1-2$ GHz) image}
Our new VLA L-band ($1-2$ GHz) C configuration image shows evidence of extended diffuse emission surrounding the jetted structures (see the 3$\sigma$ contour in the bottom right panel of figure \ref{nouvelle_fig2}). This structure has a full extent of $570$ kpc along the direction of the jets and extends beyond the original radio jetted emission reported in \cite{mcnamara_heating_2005}, \cite{cohen_resolving_2005}, and \cite{mcnamara_energetic_2009}. It also appears to be consistent with the wider radio emission reported in \cite{biava_constraining_2021}.

The extended diffuse emission that we detect could thus be the $1.5$ GHz counterpart of the jet-related emission detected in \cite{biava_constraining_2021}. We note that the structure we detect with the VLA observations seems to be somewhat spatially linked to the jets (its extent being larger along the direction of the jets compared to the orthogonal direction). However, \cite{giacintucci_occurrence_2017} showed that essentially all massive strong cool core clusters such as MS0735 should host a radio mini-halo. The extended diffuse emission could thus also be related to a new radio structure, such as a radio mini-halo. The size of the structure is consistent with the size of a radio mini-halo which are typically on the order of \textasciitilde $100-500$ kpc (\citealt{gitti_radio_2016}). We explore these two possibilities further in section \ref{Properties_diffuse_emission}. 


We also note the presence of an important quantity of point sources in our VLA L-band ($1-2$ GHz) C configuration image (see figure \ref{nouvelle_fig2} bottom right panel) as is expected due to the good spatial resolution (beam size of $13.5$ $ \times $ $19.2$ arcsec) and large field of view of the VLA at such low-frequencies. Using the compact source image presented in figure \ref{PS}, we computed the total number of point sources with a flux density above $3\sigma_{\text{rms}}$ in a region of radius equal to $4.76$ arcmin ($1$ Mpc at the cluster's redshift) from the central radio source, and found a total of 38 point sources. We note that we did not detect any extended sources. We also computed the number of point sources in regions of the same radius located outside of the cluster. In order to do that, we identified four regions located at $19.04$ arcmin ($4$ Mpc at the cluster's redshift) from the center of MS0735 and respectively located at the northeast, northwest, southeast and southwest of the center. In each of these regions of radius equal to $4.76$ arcmin ($1$ Mpc at the cluster's redshift), we computed the number of point sources and found quantities of 19, 9, 15 and 9 point sources respectively. The quantity of point sources inside the cluster is significantly higher than outside of it. We thus argue that we detect an AGN population within the cluster's region which could contribute to enhance the computed radio power (see the computation for the radio power of the P-band data in section \ref{Pbandcomputation}).

\subsubsection{VLA P-band ($224-480$ MHz) image}
\label{Pbandcomputation}
In the new VLA P-band ($224-480$ MHz) C configuration image, we detect emission out to $2$ Mpc in full extent along the direction of the jets (see the 3$\sigma$ contour of the bottom middle panel of figure \ref{nouvelle_fig2}). This is much larger than the diffuse structure detected in our VLA L-band ($1-2$ GHz) image. However, this is partially due to the two sources S1 and S2 that we detect in the VLA L-band ($1-2$ GHz) image (see right panels of figure \ref{nouvelle_fig2} where we identified S1 and S2 in yellow). The full extent of the extended structure related to MS0735 would then be better delimited by the 21$\sigma$ contour which has an extent of $\sim900$ kpc. We note that we tested various multiples of $\sigma$ to compute the extent of the extended structure and 21$\sigma$ was visually determined to be the best choice: it best encompasses the extended emission related to the center while excluding the emission related to the sources S1 and S2. We also note that since the VLA P-band ($224-480$ MHz) is at lower frequencies than the VLA L-band ($1-2$ GHz), these observations could be detecting emission from an older population of electrons that are not detected in the VLA L-band ($1-2$ GHz). 

To further investigate the origin of this extended emission, we compare the VLA P-band ($224-480$ MHz) C configuration image obtained with a robust parameter of 2.0 to the previously published VLA P-band ($327$ MHz) A configuration image in figure \ref{comparison_old_new_Pband}. We used the VLA P-band ($224-480$ MHz) C configuration image obtained with a robust parameter of 2.0 in order to maximize the detection of diffuse large scale emission. We then computed the radio power of both images as follows. First, we computed the total flux density included in the $3\sigma_\text{rms}$ contour region (where $\sigma_\text{rms} = 1.3$ mJy beam$^{-1}$ is the rms noise in the vicinity of the jets) for the previously published VLA P-band ($327$ MHz) A configuration image to have an estimate of the flux density attributable to the central compact source and to the jets. We used the CASA task \texttt{IMSTAT} to compute the flux density and we used the following expression to estimate $\sigma_\text{S}$ the uncertainty on the flux density (see \citealt{cassano_revisiting_2013}):
\begin{equation}
    \sigma_\text{S} = \sqrt{(\sigma_\text{cal} S)^2 + (\sigma_\text{rms} \sqrt{N_\text{beam}})^2 + \sigma_\text{sub} ^2}
\end{equation}

where $\sigma_\text{cal}$ is the uncertainty in the calibration of the absolute flux density scale which is typically between 5-8 per cent (we used 8 per cent as a conservative value), $S$ is the flux density (we note that approximately $7$ per cent of the computed flux density is attributable to the central point source), $\sigma_\text{rms}$ is the rms noise of the image, $N_\text{beam}$ is the number of beams covering the region where the flux density is computed and $\sigma_\text{sub}$ is the uncertainty due to the error in the source subtraction (not applicable here -- no point source subtraction was performed). We find a total flux density of $S_{327\text{MHz, A}} = (0.60\pm0.05)$ Jy for the previously published VLA P-band (327 MHz) A configuration.

From the flux density measurement, we computed the associated k-corrected radio power $P$ using the following expression (see \citealt{van_weeren_distant_2014}):

\begin{equation}
    P = 4\pi S D_L^2 (1+z)^{-(-\alpha + 1)} 
\end{equation}
where $D_L$ is the luminosity distance, $z$ is the redshift and $\alpha$ is the spectral index. We used a spectral index range that is typical for diffuse radio emission such as past jetted activity or radio mini-halos of $\alpha = (1.5 \pm 0.4)$ and we found a total radio power of $P_{327\text{MHz, A}} = (1.10\pm 0.06)\times 10^{26}$ WHz$^{-1}$ for the previously published VLA P-band ($327$ MHz) A configuration image.

We then proceeded to compute the flux density and the k-corrected radio power for our new VLA P-band ($224-480$ MHz) C configuration image using a region including the emission above 21$\sigma_\text{rms}$ (where $\sigma_\text{rms} = 0.5$ mJy beam$^{-1}$ is the rms noise in the vicinity of the cluster). This corresponds to the smaller contour shown on figure \ref{comparison_old_new_Pband} left panel and we decided to use this region to compute the flux in order to ensure that the flux density computation is not contaminated by the sources S1 and S2. We find that the total flux density is $S_{224-480\text{MHz, C}} = (0.70\pm0.06)$ Jy. Using the same typical spectral index of $\alpha = (1.5 \pm 0.4)$, the computed k-corrected radio power is $P_{224-480\text{MHz, C}} = (1.29\pm 0.08)\times 10^{26}$ WHz$^{-1}$.

By computing the difference between these radio powers, we find an excess radio power of $P_\text{excess, 224-480 MHz} = (2\pm 1)\times 10^{25}$ WHz$^{-1}$ that we associate with the diffuse emission. 

We also computed the difference in radio power using the new VLA P-band ($224-480$ MHz) C configuration image obtained with a robust parameter of -2.0 (see top middle panel of figure \ref{nouvelle_fig2}) and found the same excess radio power. This is reassuring considering that the VLA P-band ($224-480$ MHz) C configuration image obtained with a robust parameter of -2.0 does not have extended emission overlapping with the emission of the sources S1 and S2. We note that this radio power difference has a very large relative uncertainty of $50$ per cent. Nonetheless, this value confirms that the new VLA images detect a diffuse radio structure centered on the cluster that was not detected as part of the \cite{mcnamara_energetic_2009} study. We note that additional diffuse emission is expected due to 
the use of the C configuration instead of the A configuration. Indeed, the C configuration can detect structures up to scales on the order of
$67$ arcmin (or $14$ Mpc at the cluster's redshift) instead of on the order of $2.5$ arcmin (or $525$ kpc at the cluster's redshift) for the A configuration. 

\subsection{The nature of the newly-discovered VLA extended radio emission}
\label{Properties_diffuse_emission}

As was pointed out in the previous section, we found evidence of additional diffuse emission surrounding the jetted emission previously detected in this system (e.g. \citealt{mcnamara_energetic_2009}). However, it is not yet clear if this newly detected emission is a diffuse counterpart to the jet-related emission presented in \cite{biava_constraining_2021}, if it is related to a new structure such as a radio mini-halo, or if it is a combination of both of these hypotheses. In the next two subsections, we investigate these two hypotheses independently. 

\subsubsection{Jet-related counterpart to the 144 MHz LOFAR data}

To test the hypothesis of a jet-related counterpart to the $144$ MHz LOFAR data, we used the spectral index to scale the flux density of the jet-related emission detected by LOFAR (see \citealt{biava_constraining_2021}). We scaled it to $352$ MHz to be able to directly compare its flux density to the flux density of our detection. The total flux density of the lobes plus core at 144 MHz is $S_{144\text{MHz}} = (4.7\pm0.5)$ Jy. In order to scale it to $352$ MHz, we used a spectral index of $\alpha = (2.3 \pm 0.3)$ which is an average value between the average spectral index in their $144$ - $235$ MHz spectral index map and in their $144$ - $610$ MHz spectral index map. We find a flux density of $S_{352\text{MHz}} = (0.62\pm0.05)$ Jy, which is comparable to the VLA P-band ($327$ MHz) A configuration flux density of $S_{327\text{MHz, A}} = (0.60\pm0.05)$ Jy. The scaled value is smaller than the flux density of our new VLA P-band ($224-480$ MHz) C configuration image of $S_{224-480\text{MHz, C}} = (0.70\pm0.06)$ Jy, which suggests that we detected slightly more diffuse emission.

The spectral index map that we derived between VLA P-band ($224-480$ MHz) and VLA L-band ($1-2$ GHz) (see figure \ref{SIM}) shows values comparable to the spectral index maps presented in \cite{biava_constraining_2021}. Our spectral index map shows a gradient going from a value of $\alpha_{\text{center}} = 2.0$ at the center to a value of approximately $\alpha_{\text{outskirt}} = 3.0$ in the outskirt. The central value of $\alpha_{\text{center}} = 2.0$ is most likely the result of the superposition of the lobes' steep emission and the AGN flat emission resulting from the poor spatial resolution (\citealt{hogan_comprehensive_2015}) or a projection effect. The outskirt's steep spectral index agrees with the spectral index map between $610$ MHz and $1.42$ GHz presented in \cite{biava_constraining_2021} for which the outskirt spectral index was $\alpha_{610}^{1420} = 3.4 \pm 0.6$. 
This result strengthens the hypothesis that the outskirt emission is associated with an older population of accelerated electrons. Moreover, as was previously pointed out in \cite{vantyghem_cycling_2014} and in \cite{biava_constraining_2021}, MS0735 hosts multiple generations of outbursts. It was thus expected that the central part of the cluster should exhibit flatter spectral indexes than the outer parts due to a more recent re-acceleration of the electron population by a second or third generation of outbursts from the central AGN. We however argue that, due to the poor spatial resolution of the map, we cannot distinguish the complex gradient in spectral index associated with the multiple generations of outbursts.


\subsubsection{Radio mini-halo candidate}

To test the hypothesis of a radio mini-halo candidate detection, we scaled the radio power in excess of our new VLA P-band ($224-480$ MHz) C configuration image in comparison with the previously published VLA P-band ($327$ MHz) A configuration image of \cite{mcnamara_energetic_2009} to $1.4$ GHz. To do that scaling, we used a typical spectral index for radio mini-halos of $\alpha = 1\pm 0.2$ (\citealt{giacintucci_expanding_2019}). We find a radio power of $P_\text{excess, 1.4 GHz} = (4 \pm 2) \times 10^{24}$ WHz$^{-1}$. 

By comparing this radio power to the radio mini-halo detections reported in \cite{giacintucci_expanding_2019}, we see that this radio power is fully consistent with typical radio mini-halos associated with massive cool core clusters. In \cite{giacintucci_expanding_2019}, the reported radio mini-halos have $1.4$ GHz radio powers ranging between $(0.06\pm0.01) \times 10^{24}$ WHz$^{-1}$ for 2A 0335+096 and $(27\pm2) \times 10^{24}$ WHz$^{-1}$ for RX J1347.5--1145. We note that the reported radio power for  RX J1347.5--1145 is much larger than the second most luminous radio mini-halo for which the radio power is not extrapolated from a different frequency flux measurement: RX J1720.1+2638 has a radio power of $(5.3\pm0.3) \times 10^{24}$ WHz$^{-1}$. This radio power is comparable to the value we find for MS0735. The radio power of MS0735 is also consistent with the mini-halos listed in \cite{richard-laferriere_relation_2020}. If the newly detected emission is indeed a radio mini-halo, it would rank among the most luminous radio mini-halos.

In figure \ref{correlations}, we present two radio mini-halo correlations that were previously established: the correlation between the radio mini-halo radio power ($P_{1.4\text{GHz}}$) and the X-ray cavity power, and the correlation between the radio mini-halo radio power ($P_{1.4\text{GHz}}$) and the X-ray luminosity inside a radius of $600$ kpc ($L_X$) (see \citealt{richard-laferriere_relation_2020}). The linear fits in log-log space were done using the BCES-bisector and BCES-orthogonal regression algorithms (\citealt{akritas_linear_1996}; see \citealt{nemmen_universal_2012} for one example of application of the BCES method) on the radio mini-halos listed in \cite{richard-laferriere_relation_2020}. As identified in \cite{richard-laferriere_relation_2020}, the candidate and uncertain radio mini-halos (in blue in figure \ref{correlations}) are radio mini-halos for which the morphology is unusual or for which previous measurements were not deep enough to conclude confidently on their nature. We added to the correlations a purple star corresponding to the properties of the extended radio emission detected in MS0735 in the current paper. Note that we assumed that the excess flux is only attributable to a mini-halo structure when we computed the radio power for the purple star (see section \ref{combination}). The radio power of the extended radio emission in MS0735 is among the upper part of the radio mini-halos, like its cavity power, while its X-ray luminosity is among the lower part of the listed radio mini-halos. MS0735 fits relatively well on these correlations. Moreover, we note that MS0735 also falls well within another trend from \cite{richard-laferriere_relation_2020}, the radio power - cluster's mass trend, although the scatter for that trend is large.

Given the location of the newly discovered emission, its radio power, and the expected presence of a radio mini-halo in MS0735 (MS0735 being a massive cool-core cluster), we argue that this newly detected diffuse emission could well be related to a radio mini-halo. Moreover, it fits well with the correlations presented in figure \ref{correlations}. However, due to the limited spatial resolution of our observations and 
the elongated shape of the structure, we cannot 
exclude the possibility that it is jet-related.  

\subsubsection{Combined emission from jet-related emission and mini-halo}
\label{combination}
Our observations do not allow us to determine with certainty if the emission is jet-related or associated with a diffuse structure like a mini-halo, or if it is attributable to a combination of both. We would need high-resolution observations of this extended part to determine if it is linked with the jets.

\section{Concluding remarks}

We have detected additional diffuse radio emission associated with the strong cool core massive galaxy cluster MS0735 using 5 hours of VLA P-band ($224-480$ MHz) and 5 hours of VLA L-band ($1-2$ GHz) radio data. 

The detected structure has a radio power scaled at $1.4$ GHz of $P_\text{excess, 1.4 GHz} = (4 \pm 2) \times 10^{24}$ WHz$^{-1}$. It has a total extent of $\sim900$ kpc along the direction of the radio jets and of $\sim500$ kpc along the direction perpendicular to the radio jets, as seen with the VLA P-band ($224-480$ MHz) C configuration data. 

The nature of this newly discovered diffuse component remains however unclear. It could be associated with radio plasma that has diffused out of the X-ray cavities, or it could represent a different type of radio structure such as a radio mini-halo, although these two phenomena are most likely closely linked. Indeed, MS0735 is a massive strong cool core cluster and should therefore host a radio mini-halo (\citealt{giacintucci_occurrence_2017}). The radio power scaled at $1.4$ GHz of $P_\text{excess, 1.4 GHz} = (4 \pm 2) \times 10^{24}$ WHz$^{-1}$ and the size of the structure are consistent with those of a radio mini-halo. The cluster also exhibits mass, cavity power, X-ray luminosity, and radio power at 1.4 GHz consistent with previously derived radio mini-halo correlations. However, observations at higher spatial resolution are needed to fully characterize the properties and nature of this component. 

Another possibility is that we may be witnessing the early stages of formation of a radio mini-halo. If they originate or are directly linked to the jetted activity of the central AGN, as argued by deep VLA images of the Perseus cluster (\citealt{gendron-marsolais_deep_2017}) and the correlations found by \cite{richard-laferriere_relation_2020}, radio mini-halos may then form out of the jetted activity, for example by diffusing out of X-ray cavities and then being re-accelerated by continuous AGN activity as has been observed in MS0735 (\citealt{vantyghem_cycling_2014}).

Overall, our study showcases the importance of having a good combination of deep observations and high spatial resolution, especially when investigating a cluster with such large radio jets. The Next Generation Very Large Array (ngVLA) or the Square Kilometre Array (SKA) will be ideal instruments to probe systems like MS0735 by combining very high sensitivity to flux density and spatial resolution. The SKA phase 2 is also expected to be able to detect up to \textasciitilde 1900 radio mini-halos at redshifts $z < 0.6$ (\citealt{gitti_ska_2014}).

\section*{Acknowledgements}
We thank Laura Bîrzan and Brian McNamara for supplying the $327$ MHz A configuration VLA image that was presented in \cite{mcnamara_energetic_2009} and was processed as part of \cite{birzan_radiative_2010}. We also thank the Scientific Editor and the anonymous referee for providing helpful comments.

T.B. acknowledges financial support from the physics department of the Université de Montréal. J.H.-L.  acknowledges  support  from the Natural
Sciences and Engineering Research Council of Canada (NSERC) via the Discovery grant program, as well as the Canada Research  Chair  program.

The National Radio Astronomy Observatory is a facility of the National Science Foundation operated under cooperative agreement by Associated Universities, Inc.

\section*{Data availability}
The radio data are available in the NRAO Science Data Archive at \url{https://archive.nrao.edu/archive/advquery.jsp} and can be accessed using the project number 18B-171.

The X-ray data are available in the Chandra Data Archive at \url{https://cda.harvard.edu/chaser/} and can be accessed using the Observation IDs 10468, 10469, 10470, 10471, 10822, 10918 and 10922.



\bibliographystyle{mnras}
\bibliography{references} 





\bsp	
\label{lastpage}
\end{document}